\documentclass[letterpaper, 10pt, conference]{ieeeconf}
\IEEEoverridecommandlockouts
\usepackage{amsmath}
\usepackage{amssymb}
\usepackage{amsthm}
\usepackage{xcolor}
\usepackage{graphicx}
\usepackage{algorithm}
\usepackage{algpseudocode}
\usepackage{tikz}
\usepackage{comment}
\usetikzlibrary{positioning, shapes, arrows, calc, decorations.markings}
\tikzstyle{block} = [draw, rectangle, line width=0.5mm, inner sep = 1mm, minimum width = 2.5em]
\tikzstyle{sum} = [draw, circle, line width=0.5mm, minimum width=3mm, inner sep=0mm]
\tikzstyle{sig} = [draw, ellipse, line width=0.5mm, minimum width=9mm, inner sep = 1mm]
\newcommand{\jnote}[1]{{\color{blue} Jing: #1}}
\newcommand{\onote}[1]{{\color{red} Olle: #1}}

\newcommand{\bmat}[1]{\begin{bmatrix} #1 \end{bmatrix}}

\newcommand{\Phisf}{\mathbf \Phi^\text{SF}}
\newcommand{\tPhisf}{{\mathbf \Phi}^\text{SF}}
\newcommand{\Phifc}{\mathbf \Phi^\text{KF}}
\newcommand{\tPhifc}{{\mathbf \Phi}^\text{KF}}
\newcommand{\bPhi}{\mathbf \Phi}
\newcommand{\hbPhi}{\widehat{\mathbf \Phi}}
\newcommand{\R}{\mathbb R}

\newcommand{\balpha}{\boldsymbol \alpha}
\newcommand{\bbeta}{\boldsymbol \beta}
\newcommand{\bgamma}{\boldsymbol \gamma}
\newcommand{\btheta}{\boldsymbol \theta}
\newcommand{\bzeta}{\boldsymbol \zeta}
\newcommand{\bdelta}{\boldsymbol \delta}
\newcommand{\bu}{\mathbf u}

\newcommand{\bwhat}{\widehat{\mathbf w}}
\newcommand{\bwtilde}{\widetilde{\mathbf w}}

\newcommand{\bvhat}{\widehat{\mathbf v}}
\newcommand{\bvtilde}{\widetilde{\mathbf v}}
\newcommand{\by}{\mathbf y}
\newcommand{\Nin}{\mathcal N_\text{in}}
\newcommand{\Nout}{\mathcal N_\text{out}}
\newcommand{\s}{\mathcal{S}}
\newcommand{\norm}[1]{\left\Vert#1\right\Vert}

\newcommand{\xu}{\bmat{\mathbf{x}\\ \mathbf{u}}}
\newcommand{\wv}{\bmat{\mathbf{w}\\ \mathbf{v}}}
\newcommand{\PHI}{\bmat{\bPhi_{xx} & \bPhi_{xy}\\ \bPhi_{ux} & \bPhi_{uy}}}

\newcommand{\phixi}{\varphi_x}
\newcommand{\phiui}{\varphi_u}
\newcommand{\bphixi}{\boldsymbol{\phixi}}
\newcommand{\bphiui}{\boldsymbol{\phiui}}

\newcommand{\tran}{{\mkern-1.5mu\mathsf{T}}}
\newcommand{\ssreal}[4]{%
    \left[\begin{array}{c| c}
        #1 & #2 \\
        \hline
        #3 & #4
    \end{array}
    \right]
}
\theoremstyle{definition}
\newtheorem{definition}{Definition}

\newtheorem{theorem}{Theorem}

\newfloat{Subroutine}{htbp}{Subroutine}
\algnewcommand\algorithmicinput{\textbf{Input:}}
\algnewcommand\algorithmicoutput{\textbf{Output:}}
\algnewcommand\Input{\item[\algorithmicinput]}%
\algnewcommand\Output{\item[\algorithmicoutput]}%

\title{\LARGE \bf
On Infinite-horizon System Level Synthesis Problems
}

\author{ Olle Kjellqvist$^*$ and Jing Yu$^*$
\thanks{* The authors contributed equally to this work.}
\thanks{OK is with the Department of Automatic Control, Lund University,
        Lund, Sweden
        {\tt\small olle.kjellqvist@control.lth.se}, and has received funding from the European Research Council (ERC) under the European Union’s Horizon 2020 research and innovation programme under grant agreement No 834142
(ScalableControl).}%
\thanks{JY is with the Department of Computing and Mathematical Sciences, California Institute of Technology, Pasadena, CA, USA,
        {\tt\small jing@caltech.edu}}%
}

\begin{document}

\maketitle
\thispagestyle{empty}
\pagestyle{empty}

\begin{abstract}
    System level synthesis is a promising approach that formulates structured optimal controller synthesis problems as convex problems.
    This work solves the distributed linear-quadratic regulator problem under communication constraints directly in infinite-dimensional space, without the finite-impulse response relaxation common in related work.
    Our method can also be used to construct optimal distributed Kalman filters with limited information exchange.
    We combine the distributed Kalman filter with state-feedback control to perform localized LQG control with communication constraints. We provide agent-level implementation details for the resulting output-feedback state-space controller.
\end{abstract}

\section{INTRODUCTION}

In recent years, control design for networked dynamical systems has seen tremendous interest and progress.
An important problems is to impose structures on the controllers, such as sparsity for distributed or localized control \cite{lessard2012optimal, wang2014localized}, and communication delay constraints \cite{wang2014localized, feyzmahdavian2012optimal}. Such control design problems are challenging due to the non-convex nature of the problem \cite{rotkowitz2005characterization}. 

Lately, researchers have focused on novel controller parameterization that admits convex formulation \cite{wang2019system, furieri2019input, sabuau2021network}, with System Level Synthesis (SLS) emerging as a promising and unified framework for structured controller synthesis \cite{anderson2019system}. A vital feature of the SLS framework is that both the synthesis and the implementation of the structured controller can be done \textit{locally}, thus scaling favorably with the number of subsystems in a network. 


All current SLS-based control methods require both the parameterization and implementation to have finite impulse responses (FIR), with the exception of \cite{yu2021localized, fisher2022system}.
This is because
{optimal controller synthesis, despite the choice of convex reparameterization, is an infinite-dimensional non-convex optimization problem over dynamical systems. 
The current method of choice to relax the problem into a tractable {finite-dimensional} optimization problem is to restrict the optimization variable to having a finite impulse response. Such relaxation technique is required for many parameterizations other than SLS \cite{zheng2022system}. Although previous work almost exclusively uses FIR approximations, we emphasize that FIR is not a requirement for SLS, but rather a convenient way to use off-the-shelf optimization software.} 
Therefore, lifting the FIR constraint will render SLS applicable to detectable and stabilizable systems. 

An exception is \cite{yu2021localized}, where the authors showed that a class of the infinite-horizon state-feedback SLS problem has natural connections to the Riccati solution for linear quadratic regulator (LQR) problems. The resulting SLS controller has a state-space form that significantly reduces the required memory compared to the FIR SLS controllers.

It is a natural next step to investigate the correspondence between infinite-horizon output-feedback SLS and the linear quadratic Gaussian (LQG) control \cite{kalman1960contributions, wonham1968separation}.




\textbf{Contribution.}
This paper generalizes a previous result on infinite-horizon state-feedback SLS problem, and investigates a class of infinite-horizon output-feedback SLS problems. In particular, we study an output-feedback SLS problem that corresponds to a class of LQG problems with structural constraints expressable as convex sparsity constraints under the SLS parameterization. Our contribution is three-fold.
    (1) We generalize the result on the infinite-horizon state-feedback SLS solution \cite{yu2021localized} to scenarios where communication delay constraints among subsystems in a network can be incorporated into controller synthesis and implementation. 
    (2) We provide a suboptimal solution to a class of infinite-horizon output-feedback SLS problems. Our solution leverages an analogous separation principle for SLS parameterization, where the proposed generalized infinite-horizon state-feedback SLS solution is used. 
     A key advantage of our approach is the ability to compute the parameters of each subsystem locally in one swoop using \textit{local} information without iterations or communications among subsystems, which were required by previous methods. 
    (3) We demonstrate an internally stabilizing output-feedback controller that is distributed and localized based on the proposed suboptimal solution.   
    The proposed state-space controller has a fixed, low memory requirement, unlike existing FIR-based SLS controllers where the length of the memory grows linearly with the FIR horizon.  

\textbf{Paper Structure.} 
We introduce the infinite-horizon SLS problems and related concepts in Section~\ref{sec:prelim}. Section~\ref{sec:delay} proposes a generalized approach that solves the infinite-horizon state-feedback SLS problems with communication delay and localization constraints. In Section~\ref{sec:solution}, we construct a suboptimal solution to the infinite-horizon output-feedback SLS problem using the proposed optimal state-feedback solution. In Section~\ref{sec:controller}, we show the implementation of the state-space controller associated with the constructed suboptimal solution. Section~\ref{sec:simulation} demonstrates numerical simulation that corroborates with the theoretical results.

\textbf{Notation.}
Bold font $\mathbf x$ and $\mathbf G$ denote signals $\mathbf{x} = \{x(t) \}_{t = 0}^\infty$ with $x(t) \in \R^n$, and proper transfer matrices $\mathbf{G}(z) = \sum_{i = 0}^\infty z^{-i}G[i]$ with convolution kernels $G[i]\in \R^{m \times n}$. The $j^\text{th}$ standard basis vector is denoted as $e_j$. For a matrix $A$, $A(i, j)$ refers to the $(i, j)^\text{th}$ element, $A(i, :)$ to the $i^\text{th}$ row and $A(:, j)$ to the $j^\text{th}$ column. Non-negative integers are denoted as $\mathbb{N}_{+}$. We write $A \succ B$ ($A \succeq B$) to mean that $A - B$ is a positive (semi)definite matrix. We use $\mathbb{R}\mathcal{H}_\infty$ for the space of all proper and real rational stable transfer matrices and denote $F \in \frac{1}{z}\mathbb{R}\mathcal{H}_\infty$ if and only if $zF \in \mathbb{R}\mathcal{H}_\infty$.

\section{Preliminaries}
\label{sec:prelim}
We consider the following dynamical system
\begin{equation}
    \label{eq:dynamics}
    \begin{aligned}
        x(t+1) & = Ax(t) + Bu(t) + w(t) \\
        y(t) & = Cx(t) + v(t),
    \end{aligned}
\end{equation}
where $w(t)\sim \mathcal{N}(0,W)$, and $v(t)\sim \mathcal{N}(0,V)$ are respectively state and measurement noise independently and identically drawn at each time from the zero-mean Gaussian distributions. Here $W$ and $V$ are positive definite matrices. The general control design objective is to synthesize a linear controller $\mathbf{K}$ such that control action computed as $\mathbf{u} = \mathbf{Ky}$ stabilizes the closed-loop system while optimizing over a quadratic cost $J(\mathbf{x},\mathbf{u}) := \lim_{T\rightarrow \infty} \mathbb{E} \sum_{t=0}^T  x(t)^{\tran}Rx(t) + u(t)^\tran Q u(t)$, with $Q$, $R \succ 0$ . This is known as the linear quadratic Gaussian (LQG) problem, where the optimal controller is given by the combination of a state-feedback controller and a state observer\cite{kalman1960contributions}.

In this paper, we investigate a \emph{distributed} variant of the LQG problem, where \eqref{eq:dynamics} is constructed from a network of heterogeneous subsystems with dynamical coupling. In particular,
we aim to design a stabilizing output-feedback controller that respects \emph{communication delays} and \emph{localization constraints}. For detalied discussions of such constraints, we refer interested readers to \cite{yu2021localized, wang2018separable}.


\subsection{System Level Synthesis}

The SLS theory approaches the constrained output-feedback control design problem described above by characterizing all achievable closed-loop mappings (CLMs) from $\mathbf{w},\mathbf{v}$ to $\mathbf{x}$, $\mathbf{u}$ under an internally stabilizing controller $\mathbf{K}$. Then, using any achievable CLMs, SLS provides an implementation of the controller $\mathbf{K}$ that realizes the prescribed CLMs. This is made precise in the following result. 

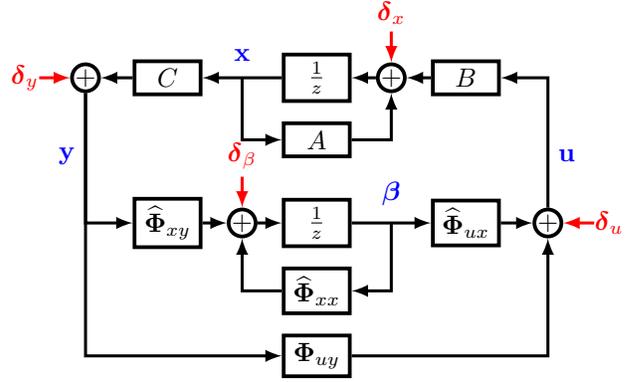
\begin{figure}
    \centering
    \scalebox{1}{\begin{tikzpicture}[node distance = 4.5em, line width=0.4mm, inner ysep=6pt, >=latex]
    \node[block] (delay1) {$\frac{1}{z}$};
    \node[block, below = .7em of delay1] (A) {$A$};
    \node[block, left = 3em of delay1] (C2) {$C$};
    \node[block, right = 3em of delay1] (B2) {$B$};
    \node[sum] at ($(delay1)!0.5!(B2)$) (sumx){$+$};
    \node[block, below = 1.5em of A] (delay2) {$\frac{1}{z}$};
    \node[block, below = .7em of delay2] (Pxx) {$\hbPhi_{xx}$};
    \node[block, left = 3em of delay2] (Pxy) {$\hbPhi_{xy}$};
    \node[block, right = 3em of delay2] (Pux) {$\hbPhi_{ux}$};
    \node[block, below = .7em of Pxx] (Puy) {$\bPhi_{uy}$};
    \node[sum] at ($(delay2)!0.5!(Pxy)$) (sumb){$+$};
    \node[sum, left = 1.2em of C2] (sumy) {$+$};
    \node[sum, right = 1.2em of Pux] (sumu) {$+$};
    
    \node[right = .8em of sumu, red] (du) {$\boldsymbol \delta_u$};
    \node[above = .8em of sumx, red] (dx) {$\boldsymbol \delta_x$};
    \node[left = .8em of sumy, red] (dy) {$\boldsymbol \delta_y$};
    \node[above = .8em of sumb, red] (db) {$\boldsymbol \delta_\beta$};
    
    \draw[->] (sumu) |- node[blue, pos = 0.2, right] {$\bu$} (B2);
    \draw[->] (B2) -- (sumx);
    \draw[->] (sumx) --  (delay1);
    \draw[->] (delay1) -- node[above, blue] {$\mathbf x$} (C2);
    \draw[->] ($(C2)!0.5!(delay1)$) |- (A);
    \draw[->] (A) -| (sumx);
    \draw[->] (C2) -- (sumy);
    \draw[->] (sumy) |- (Pxy);
    \draw[->] (Pxy) -- (sumb);
    \draw[->] (sumb) -- (delay2);
    \draw[->] (delay2) -- node[above, blue] {$\bbeta$} (Pux);
    \draw[->] (Pux) -- (sumu);
    \draw[->] ($(Pux)!0.5!(delay2)$) |- (Pxx);
    \draw[->] (Pxx) -| (sumb);
    
    \draw[->] (sumy) |- node[blue, pos = 0.12, left] {$\by$} (Puy);
    \draw[->] (Puy) -| (sumu);
    
    \draw[->, red] (du.west)++(1.3mm, 0) -- (sumu);
    \draw[->, red] (dx.south)++(0, 1.3mm) -- (sumx);
    \draw[->, red] (dy.east)++(-1.3mm, 0) -- (sumy);
    \draw[->, red] (db.south)++(0, 1.3mm) -- (sumb);
\end{tikzpicture}}
    \caption{Output-feedback controller architecture. Here, $\hbPhi_{xx} = z(I - z\bPhi_{xx})$, $\hbPhi_{ux} = z\bPhi_{ux}$ and $\hbPhi_{xy} = -z\bPhi_{xy}$. The controller is internally stable; the closed-loop maps from perturbations $(\bdelta_x, \bdelta_y, \bdelta_u, \bdelta_\beta)$ to internal signals $(\mathbf x, \by,\bu, \bbeta)$ are stable.}
    \label{fig:sls_of}
\end{figure}

\begin{theorem}[\cite{anderson2019system}]
\label{thm:sls}
Strictly proper linear CLMs $\bPhi_{xx}$, $\bPhi_{xy}$, $\bPhi_{ux}$, $\bPhi_{uy} \in \frac{1}{z} \R\mathcal{H}_\infty$ can be achieved by a linear internally stabilizing controller $\mathbf{K}$ if and only if
\begin{subequations}
\label{eq:output-feasibility}
\begin{align}
    \bmat{zI - A & -B} \PHI & = \bmat{I & 0} \label{eq:feasibility_x}\\
    \PHI \bmat{zI - A \\ -C} & = \bmat{I \\ 0}, \label{eq:feasibility_u}
\end{align}
\end{subequations}
where $\bPhi_{xx}$, $\bPhi_{xy}$, $\bPhi_{ux}$, $\bPhi_{uy}$ maps $\mathbf{w},\mathbf{v}$ to $\mathbf{x}$, $\mathbf{u}$ under a controller $\mathbf{K}$, i.e.,$\xu = \PHI \wv $. 
In particular, $\mathbf{K}$ can be implemented as the following, which is illustrated in Figure \ref{fig:sls_of}:
\begin{equation}
    \label{eq:of-controller}
    \begin{aligned}
        z\bbeta & = \hbPhi_{xx}\bbeta + \hbPhi_{xy}\by \\
        \bu & = \hbPhi_{ux}\bbeta + \bPhi_{uy}\by,
    \end{aligned}
\end{equation}
where $\hbPhi_{xx} = z(I - z\bPhi_{xx})$, $\hbPhi_{ux} = z\bPhi_{ux}$, $\hbPhi_{xy} = -z\bPhi_{xy}$, and $\beta$ is the controller internal state.
\end{theorem}
 Further, it was shown in \cite{wang2019system} that \eqref{eq:output-feasibility} is equivalent to stabilizability and detectability of \eqref{eq:dynamics}. Therefore, \eqref{eq:of-controller} parameterizes all internally stabilizing linear controller $\mathbf{K}$ for \eqref{eq:dynamics}. 

A special case of Theorem \ref{thm:sls} is when the controller is \emph{state-feedback}, i.e., $\mathbf{u} = \mathbf{K}\mathbf{x}$. In this scenario, the SLS CLMs reduce to only $\bPhi_{xw}:\mathbf{w}\rightarrow \mathbf{x}$ and $\bPhi_{uw}: \mathbf{w} \rightarrow \mathbf{u}$ with the following variation of Theorem \ref{thm:sls}.
\begin{theorem}[\cite{wang2019system}]
        \label{thrm:SLS}
For the dynamics \eqref{eq:dynamics} with $C = I$ and $v(t) \equiv 0$, CLMs $\bPhi_{xw}$ and $\bPhi_{uw}$ can be achieved by a linear internally stabilizing controller $\mathbf{K}$ if and only if 
\begin{align}
\label{eq:sf-feasibility}
    \bmat{zI - A & -B} \bmat{\bPhi_{xw}\\ \bPhi_{uw}} & = I, \quad \bPhi_{xw},\,\bPhi_{uw} \in \frac{1}{z} \R\mathcal{H}_\infty.
\end{align}
\end{theorem}

\subsection{System-level Constraints (SLCs)}
\label{sec:slc}
In this paper, we consider a networked system \eqref{eq:dynamics} composed of $N$ subsystems, where each subsystem $i$ has the following dynamics
\begin{align}
        \label{eq:local_sys}
        x_i(t+1) &= \sum_{j\in \Nin^1(i)} \left( A_{ij}x_j(t) +  B_{ij}u_j(t) \right) + w_i(t)\\
        y_i(t) &= C_i x_i(t) + v_i(t)
\end{align}
where we denote $j \in \Nin^k(i)$ if the states and control actions of subsystem $j$ affect those of subsystem $i$ in $k$ time steps through the open-loop network dynamics. Analogously, we write $i \in \Nout^k(j)$ when the states of subsystem $i$ are affected by subsystem $j$ via dynamics in $k$ time steps. 

For such systems, it is common to impose structural constraints such as localization requirement and communication delay for control design. A large class of such structural constraints can be characterized as convex constraints on CLMs. This is called the system-level constraints (SLCs) \cite{wang2019system}. Examples of SLCs include sparsity constraints on the Youla parameter, QI subspace constraints on controller $\mathbf{K}$, and FIR constraints on the closed-loop responses to noises. 

In this paper, we consider the family of SLCs that can be formulated as sparsity constraints on the CLMs. This family includes disturbance localization and communication delay constraints considered in \cite{yu2021localized}. We denote any such sparsity constraints as $\s := \{S[k]\}_{k=1}^\infty$ where each $S[k]\in \{0,1\}^{n \times n}$ are binary matrices specifying the sparsity of the kernels of CLMs. Specifically, we consider the \textit{localization SLCs}, which specifies how disturbances at each subsystem must be localized to a neighborhood.
\begin{definition}[Localization SLCs]
An SLC $\s := \{S[k]\}_{k=1}^\infty$ is called the localization SLC if $S[k]$ for all $k$ are the same binary matrices.
\end{definition}
In addition to disturbance localization which assumes instantaneous information exchange within a localization neighborhood, one may wish to incorporate \textit{communication delay SLCs}\footnote{Such delayed localization pattern corresponds to scenarios when communication delays allow disturbances to propagate through dynamics, before subsystems are able to completely attenuate and localize them. For ease of exposition, here we consider a communication delay pattern among subsystems that matches the dynamics. The results in this paper can be generalized to broader classes of communication patterns.}.  Let $\text{Sp}(\cdot)$ denote the support of a matrix.
\begin{definition}[d-Delayed localization SLCs]
For a fixed integer $d \geq 1$ and sparsity pattern of the dynamics $\mathbb{A} := \text{Sp}\left(A \right)$, a delayed localization SLC is such that $S[k] = \text{Sp}\left(\mathbb{A}^k\right)$ for $k\leq d$ and $S[k] = \text{Sp}\left(\mathbb{A}^d\right)$ for all $k \geq d$.
\end{definition}
This is sometimes called the $(A,d)$-sparsity \cite{wang2014localized} and generalizes the localization SLCs. We say $\bPhi_{xx} \in \s$ if $\Phi_{xx}[k]$ has the same sparsity pattern as $S[k]$ for all $k \in \mathbb{N}_{+}$. For the rest of paper, we consider d-delayed localization SLCs for structured controller synthesis. We assume that any given SLCs are feasible for the underlying system.

\subsection{The State-feedback and Output-feedback SLS Problem}
Given d-delayed localization SLCs $\s_x$ and $\s_u$ specifying the state and input closed-loop sparsity respectively, we now state the output-feedback (OF) SLS problem \cite{wang2019system}.
\begin{align}
 \underset{\bPhi\in \frac{1}{z} \R\mathcal{H}_\infty}{\text{min}}  &  \norm{\begin{bmatrix}Q^{1/2} & 0 \\ 0 & R^{1/2}\end{bmatrix} \PHI \begin{bmatrix}W^{1/2} & 0 \\ 0 & V^{1/2}\end{bmatrix}}_{\mathcal{H}_2} \nonumber \\
\text{s.t.} \quad &  \text{Constraints \eqref{eq:output-feasibility}}, \quad \nonumber \\
& \Phi_{xx}[k]  \,, \Phi_{xy}[k]\in  S_{x}[k]\text{ for }k\in \mathbb{N}_+ \label{eq:of-original} \tag{OF-SLS}\\
& \Phi_{ux}[k] \,,\Phi_{uy}[k] \in  S_{u}[k] \text{ for }k\in \mathbb{N}_+\,.
\nonumber
\end{align}
where we used $\bPhi$ to collectively refer to the tuple $\left(\bPhi_{xx}, \bPhi_{ux},\bPhi_{xy},\bPhi_{uy}\right)$ to reduce notation.
Control problem \eqref{eq:of-original} is similar to the classical LQG problem but differs in the  additional constraints on the system responses that corresponds to disturbance localization and information delay.

As a special case, the state-feedback (SF) SLS problem is
\begin{align}
 \underset{\bPhi_{xw},\bPhi_{uw}}{\text{min}} 
&\norm{\bmat{Q^{1/2} & 0 \\ 0 & R^{1/2}} \bmat{\bPhi_{xw} \\ \bPhi_{uw}}}_{\mathcal{H}_2}  \nonumber \\
\text{s.t.} \quad \quad \,\,\,\, &
\text{Constraints } \eqref{eq:sf-feasibility} \label{eq:sf-original} \tag{SF-SLS}\\
&\Phi_{xw}[k] \in S_x[k],\,\,\, \Phi_{uw}[k] \in S_u[k] \text{  for } k \in \mathbb{N}_+\,. \nonumber
\end{align}

\section{Optimal solution to State-feedback SLS}
\label{sec:delay}
In this section, we derive the optimal solution to \eqref{eq:sf-original}. This generalizes \cite{yu2021localized} to account for the \emph{delayed} localization constraints. This solution allows the subsequent output-feedback controller synthesis.

Because \eqref{eq:sf-original} is column-wise separable \cite{wang2018separable}, we will synthesize the closed-loop maps one column at a time so that each subsystem can synthesize the columns corresponds to its local states, in a parallel fashion. Such parallel synthesis scales favorably with the number of subsystems in the network. 
From here on, everything will be seen by the $i$th subsystem\footnote{To reduce notation, we assume each subsystem has scalar dynamics. One can alleviate this assumption by running the algorithm for multiple columns per subsystem.}. Let $\bphixi := \bPhi_{xw}(:, i)$ and $\bphiui := \bPhi_{uw}(:, i)$ with kernels $\phixi[k]$ and $\phiui[k]$ for $k \in \mathbb{N}_+$, respectively corresponding to the $i$th column of $\bPhi_{xw}$ and $\bPhi_{uw}$.
Furthermore we use $s_x[k]$ and $s_u[k]$ denote the $i{\text{th}}$ column of $S_x[k]$ and $S_u[k]$ respectively.
Each corresponding column problem to be solved locally by subsystem $i$ becomes
\begin{subequations}
\label{eq:column-sf}
\begin{align}
    \underset{\bphixi, \bphiui}{\text{min}} \quad & \sum_{k = 1}^\infty \phixi[k]^\tran Q\phixi[k] + \phiui[k]^\tran R\phiui[k] \nonumber \\
    \text{s.t.}\quad \,\,\,  & \phixi[k + 1] = A\phixi[k] + B\phiui[k]\label{eq:dynamics_extended}\\
    &\phixi[0] = 0, \quad \phixi[1] = e_i \nonumber \\
    &\phixi[k] \in s_x[k],\  \phiui[k] \in s_u[k].\label{eq:sparsity}
\end{align}
\end{subequations}
This new problem is a constrained linear quadratic optimal control problem, and would be a standard infinite-horizon LQR problem if not for the sparsity constraints~\eqref{eq:sparsity}.
We will show how to transform this problem to a finite-horizon LQR problem with time-varying dynamics.

\subsection{Derivation of the Optimal Solution}
\label{sec:column-sf}
Let $n_x[k]$ be the number of nonzero elements in the $n$-dimensional vector $s_x[k]$ and $n_u[k]$ be the number of nonzero elements in $s_u[k]$.
Then there exists a surjective matrix $M_x[k]\in \R^{(n - n_x[k])\times n}$ and an injective matrix $M_u[k]\in \R^{m \times n_u[k]}$ such that \eqref{eq:sparsity} is equivalent to

\begin{equation}
\label{eq:null}
    M_x[k]\phixi[k] = 0,\quad \phiui[k] = M_u[k]q[k],
\end{equation}
where $q[k]\in \R^{n_u[k]}$ becomes the new variable. In particular, one can construct $M_x[k]$ by horizontally stacking standard basis vectors with nonzero positions corresponding to the positions that are zero in $\phixi[k]$. On the other hand, $M_u[k]$ can be obtained similarly but with basis vectors corresponding to the nonzero positions in $\phiui[k]$.
Since $\phixi[k + 1]$ is uniquely determined by $\phixi[k]$ and $\phiui[k]$, substitution of \eqref{eq:null} into \eqref{eq:dynamics_extended} yields
\begin{equation}
\label{eq:null_next_step}
    M_x[k + 1]A\phixi[k] + \underbrace{M_x[k + 1]BM_u[k]}_{F[k]}q[k] = 0.
\end{equation}

The solutions to \eqref{eq:null_next_step} can be expressed as
\begin{equation}
\label{eq:w}
    q[k] = F[k]^\dagger M_x[k + 1]A\phixi[k] + N_F[k]r[k],
\end{equation}
where $N_F[k]\in \R^{n_u[k] \times n_r[k]}$ is a bijection onto the nullspace of $F[k]$. The vector $r[k]\in \R^{n_r[k]}$ is now our new unconstrained optimization variable.
Substituting $\phiui[k] =M_u[k]q[k] $ and \eqref{eq:w} into the optimal control problem we get the equivalent time-varying LQR problem
\begin{align}
    \underset{r[k] \in \R^{n_r[k]}}{\text{min}} \quad & \sum_{k = 1}^\infty \Big( \phixi[k]^\tran \tilde Q[k]\phixi[k] +  \nonumber\\
    & \quad 2 r[k]^\tran \tilde Z\phixi[k] + r[k]^\tran \tilde R[k] r[k] \Big) \nonumber \\
    \text{s.t.}\quad \,\,\,\,\,\,\,\, & \phixi[k + 1] = \tilde A\phixi[k] + \tilde B[k]r[k] \label{eq:inf-delayed-sf}\\
    &\phixi[0] = 0, \quad \phixi[1] = e_i,\nonumber
\end{align}
where
    \begin{align}
    \kappa[k] & = M_u[k]F[k]^\dagger M_x[k + 1]A \nonumber\\
    \tilde Z[k]  &= N_F[k]^\tran M_u[k]^\tran R \kappa[k],\quad
    \tilde Q[k]   = Q + \kappa[k]^\tran R \kappa[k]\nonumber\\
    \tilde R[k]  &= (M_u[k]N_F[k])^\tran R[k] M_u[k]N_F[k] \label{eq:data}\\
    \tilde A[k] &= A - B\kappa[k] ,\quad \tilde B[k]  =  BM_u[k]N_F[k].\nonumber
    \end{align}
Finally we note that for $k \geq d + 1$, the localization patterns are constant, implying that the dynamics matrices of the transformed problem are static for $k \geq d + 1$.
Standard dynamic programming arguments allow us to first solve the Riccati equation for the time-invariant problem for $k\geq d+1$ to get the positive definite solution $\tilde X_\star$ and the feedback gain $\tilde K_\star$, and then to solve a finite-horizon time-varying problem by replacing the cost function of each column problem \eqref{eq:inf-delayed-sf} with equivalent cost function
\begin{align}
    J &= \sum_{k = 1}^d \Big( \phixi[k]^\tran \tilde Q[k]\phixi[k]+ 
    2 r[k]^\tran \tilde Z\phixi[k] + r[k]^\tran \tilde R[k] r[k] \Big) \nonumber \\
    &+ \phixi[d+1]^\tran \tilde X_\star\phixi[d+1]. \label{eq:cost}
\end{align}


To obtain $\tilde X_\star$ and $\tilde K_\star$, we invoke the results in \cite{yu2021localized} for the time-invariant problem for with staic localization pattern.
Finally, the solution to the time-varying finite-horizon problem \eqref{eq:inf-delayed-sf} with cost \eqref{eq:cost} is given by the Riccati iteration with $\tilde X[d+1] = \tilde X_\star$,  and for $k = 1, \dots, d$,
\begin{align}
        \tilde X[k] &=\tilde Q[k] + \tilde A[k]^\tran \tilde X[k + 1] \tilde A[k]  -\nonumber \\
        &\left(\tilde A[k]^\tran \tilde X[k+1] \tilde B[k] + \tilde Z[k]\right) \nonumber\\
        &\cdot \left(\tilde R[k] + \tilde B[k]^\tran \tilde X[k + 1]B[k]\right)^{-1}
        \left(\tilde B[k]^\tran \tilde X[k + 1]\tilde A[k] \right) \nonumber \\
            \tilde K[k] &= \left(\tilde R[k] + \tilde B[k]\tilde X[k + 1]\tilde B[k]\right)^{-1} \nonumber\\
    &\quad \cdot\left(\tilde B[k]^\tran \tilde X[k + 1]\tilde A[k] + \tilde Z[k]^\tran \right). \label{eq:K}
\end{align}
Substituting $r[k] = \tilde K[k]\phixi[k]$ into \eqref{eq:w} and further into \eqref{eq:null_next_step}, one can obtain the solution to the original problem \eqref{eq:column-sf}.
We formally state the optimality of the proposed solution.
\begin{theorem}
\label{thm:optimality}
Fix delayed localization SLCs $\s_x$ and $\s_u$. The optimal solution to the infinite-horizon state-feedback SLS problem in \eqref{eq:sf-original} is given, in a column-wise fashion, by
\begin{align}
  \bPhi_{xw}^\star(:,i) = \ssreal{A_{\text{SF}}^{i}}{B_{\text{SF}}^{i}}{C_{\text{SF}}^i}{0}  ,\,\,  \bPhi_{uw}^\star(:,i) = \ssreal{A_{\text{SF}}^{i}}{B_{\text{SF}}^{i}}{K_{\text{SF}}^i}{0}   \, ,\label{eq:ss-solution}
\end{align}
where 
\begin{align}
A_{\text{SF}}^{i} &= \bmat{0 &\dots & & & 0\\ \tilde A_{\text{CL},i}[1]& 0 &\dots & & 0\\ 0& \tilde A_{\text{CL},i}[2] & 0 & \dots &0 \\ 0 & 0&\ddots & &\vdots\\ & & & \tilde A_{\text{CL},i}[d]& \tilde A_{\text{CL},\star,i} } \nonumber\\
B_{\text{SF}}^{i} &= \bmat{e_i^\tran & 0 \dots 0  }^\tran \quad
C_{\text{SF}}^{i} = \bmat{I&I\dots&I} \nonumber\\
K_{\text{SF}}^{i} &= \bmat{\tilde{K}_i[1]&\tilde{K}_i[2]& \dots&\tilde{K}_i[d]&\tilde{K}_{\star,i}}, \label{eq:i_SF}
\end{align}
with $\tilde{K}_i[k]$ and $\tilde A_{\text{CL},i}[k] := \tilde{A}[k]-\tilde{B}[k]\tilde{K}[k]$ computed using \eqref{eq:data} and \eqref{eq:K} for the $i{\text{th}}$ column problem. Matrix $\tilde A_{\text{CL},\star,i}$ and $\tilde{K}_{\star,i}$ are given by the infinite-horizon solution with static localization SLCs $\s_x \equiv S_x[d]$ and $\s_u \equiv S_u[d]$ using the method in \cite[Section IV.A]{yu2021localized}.



\end{theorem}
\begin{proof}
The optimality follows directly from the column separable property of \eqref{eq:sf-original}, and the equivalent transformations between \eqref{eq:column-sf} and \eqref{eq:inf-delayed-sf}. The finite-horizon LQR problem with cost \eqref{eq:cost} is equivalent to \eqref{eq:inf-delayed-sf} by Bellman's optimality principle. It is straightforward to verify that \eqref{eq:ss-solution} is a state-space realization of the solution to \eqref{eq:inf-delayed-sf} by substituting the optimal solution $ r[k]$ via \eqref{eq:K} into \eqref{eq:w}. 
\end{proof}
Compared to \cite{yu2021localized}, the state-space realization of the optimal CLMs given by our approach has a higher order because of the first $d$-delay pattern. If we let the first $d$ delay sparsity pattern to be the same as the localization pattern, then our approach subsumes the results in \cite{yu2021localized}.

Given the column-wise state-space description of the optimal CLMs $\bPhi_{xw}^\star(:,i)$ and $\bPhi_{xw}^\star(:,i)$, we can adopt the same state-space agent-level controller proposed in \cite[Section IV.B]{yu2021localized} by simply replacing the state-space implementation of the CLMs with the ones presented here.

\subsection{Structured Kalman Filter Design}
Theorem \ref{thm:optimality} can be used to solve the dual problem of optimal structured Kalman filter design with delayed localization SLCs for \eqref{eq:dynamics} \cite{wang2015localized}. In particular, the optimal structured infinite-horizon CLMs that map $\mathbf{w}$ and $\mathbf{v}$ to state estimation error $\mathbf{e}$ under a linear observer $\mathbf{L}$ with respect to the mean estimation error is given by the solution to the dual problem of \eqref{eq:sf-original} as shown below:
\begin{align}
\label{eq:kf-sls}
& \underset{\bPhi_{ew},\bPhi_{ev} \in \frac{1}{z} \R\mathcal{H}_\infty}{\text{min}} 
& & \norm{\bmat{W^{1/2} & 0 \\ 0 & V^{1/2}} \bmat{\bPhi_{ew}^\top \\ \bPhi_{ev}^\top}}_{\mathcal{H}_2}  \\
& \text{subject to}
& & \bmat{\bPhi_{ew} & \bPhi_{ev}} \bmat{zI - A \\ -C}= I \label{eq:kf-feasibility}\\
&&& \bPhi_{ew} \in \s_x, \quad \bPhi_{ev} \in \s_u. \nonumber
\end{align}
Readers are referred to \cite{wang2015localized} for detailed derivation. We highlight the resemblance between constraints \eqref{eq:sf-feasibility} and \eqref{eq:kf-feasibility}, and \eqref{eq:output-feasibility}. In what follows, we will use the optimal solutions from the state-feedback and Kalman-filter SLS problem to construct a suboptimal solution to the output-feedback SLS problem.

\section{A solution Inspired by Separation Principle}
\label{sec:solution}

It is well known that for a linear system, observer-based feedback is always stabilizing if the observer error dynamics are stable and the feedback gain stabilizes the state-feedback case. In \cite{wang2019system}, the authors pointed out that a similar property holds for CLMs from state-feedback and kalman-filter SLS problems described in Section~\ref{sec:delay}. 
\begin{theorem}[\cite{anderson2019system}]
\label{thm:observer_feedback}
Assume there exist stable and strictly proper transfer matrices $\Phisf = (\Phisf_{xw},\Phisf_{uw})$ and $\Phifc = (\Phifc_{ew}, \Phifc_{ev})$ satisfying
\begin{align*}
    \bmat{zI - A & -B}\bmat{\Phisf_{xw} \\ \Phisf_{uw}} & = I, \\
    \bmat{\Phifc_{ew} & \Phifc_{ev}}\bmat{zI - A \\ -C} & = I.
\end{align*}
The transfer functions
\begin{equation}\label{eq:certainty_equivalence}
\begin{aligned}
    \bPhi_{xx} & = \Phisf_{xw} + \Phifc_{ew} - \Phisf_{xw}(zI - A)\Phifc_{ew} \\
    \bPhi_{ux} & = \Phisf_{uw} - \Phisf_{uw}(zI - A)\Phifc_{ew}\\
    \bPhi_{xy} & = \Phifc_{ev} - \Phisf_{xw}(zI - A)\Phifc_{ev} \\
    \bPhi_{uy} & = - \Phisf_{uw}(zI - A)\Phifc_{ev}
\end{aligned}
\end{equation}
are strictly proper and satisfy \eqref{eq:output-feasibility}.
\end{theorem}

Then, output-feedback controller \eqref{eq:of-controller} can be constructed using CLMs from \eqref{eq:certainty_equivalence} to stabilize \eqref{eq:dynamics} while respecting the prescribed localization and communication delay constraints.


\section{Local Controller implementation}
\label{sec:controller}
This section describes Algorithm~\ref{alg:full}, which summarizes the local implementation of the global controller \eqref{eq:of-controller} in Fig.~\ref{fig:sls_of} using the localized state-feedback controllers and Kalman filters of Section~\ref{sec:delay} and Theorem~\ref{thm:observer_feedback}. 

\begin{figure*}
    \centering
    \includegraphics[width=\textwidth]{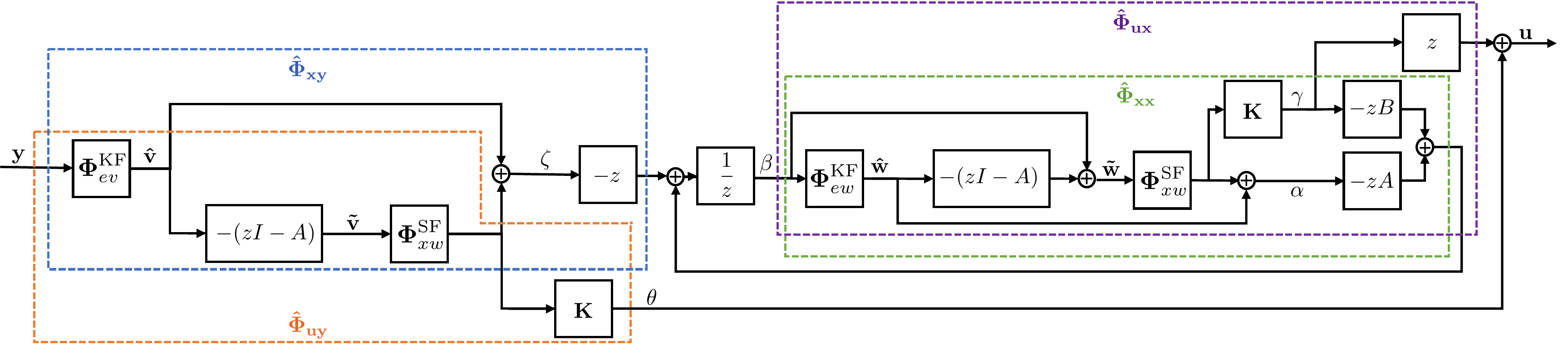}
    \caption{Controller implementation of Fig.~\ref{fig:sls_of} after plugging \eqref{eq:certainty_equivalence} in the controller \eqref{eq:of-controller}.}
    \label{fig:local_of}
\end{figure*}

\emph{Globally}, the controller after plugging \eqref{eq:certainty_equivalence} in the controller \eqref{eq:of-controller} is shown in Fig.~\ref{fig:local_of}. 
Consider the intermediate signals in Fig.~\ref{fig:local_of},
\begin{equation}
    \label{eq:greek_letters}
    \begin{aligned}
    \bzeta & =  - \Phisf_{xw}\overbrace{(zI - A)\Phifc_{ev}\by}^{\bvtilde} + \overbrace{\Phifc_{ev}\by}^{\bvhat} \\
     \btheta & = - \Phisf_{uw}(zI - A)\Phifc_{ev}\by\\
     \balpha & = \Phisf_{xw}\overbrace{(\bbeta - (zI - A)\Phifc_{ew}\bbeta)}^{\bwtilde} + \overbrace{\Phifc_{ew}\bbeta}^{\bwhat} , \\
     \bgamma & = \Phisf_{uw}(\bbeta - (zI - A)\Phifc_{ew}\bbeta). 
     \end{aligned}
\end{equation}
With these intermediate signals, we can compute the controller internal state $\bbeta$ and the control signal $\bu$ in Fig.~\ref{fig:sls_of} from $z\bbeta = -z(A\balpha + B_2\bgamma) - z\bzeta$, and $ \bu = z\bgamma + \btheta$. 

\emph{Locally}, due to the communication constraints specified in Section~\ref{sec:slc}, one can not carry out the computation described above in a centralized way. In particular, the local computation of each signal in \eqref{eq:greek_letters} involves delayed and locally available information. We now describe the information exchange among subsystems and how they compute \eqref{eq:greek_letters}. Recall that the state-feedback solution $\Phisf$ and Kalman-filters $\Phifc$ from \eqref{eq:ss-solution} are synthesized to respect the communication constraints expressed as d-Delayed localization SLCs. Denote
\begin{equation}\label{eq:phw}
\renewcommand\arraystretch{1.4}
    \Phisf_{w}(:, i) = \ssreal{A_{\text{SF}}^{i}}{B_{\text{SF}}^{i}}{I}{0}.
\end{equation}
Then $\Phisf_{xw}(:, i) = C^i_\text{SF}\Phisf_{w}(:, i)$ and $\Phisf_{uw}(:, i) = K^i_\text{SF}\Phisf_{w}(:, i)$ where $C^i_\text{SF}$ and $K^i_\text{SF}$ are from \eqref{eq:i_SF}. Computing the local components of $\balpha$ and $\bbeta$ requires only one realization of $\Phisf_{w}$ as they can share the same copy of the states within each subsystem. An analogous statement holds true for $\bzeta$ and $\btheta$. Denote the two realizations of \eqref{eq:phw} as $\Phisf_{w, \alpha}$ and $\Phisf_{w, \zeta}$. During each time step $t$, every node observes its local output $y_i(t)$ and goes through four stages of computation and communication with its neighbors leading to an update to the internal controller states and the application of the actuator signal $u_i(t)$. This is summarized in Algorithm~\ref{alg:full} with subroutines~\ref{sub:1}--\ref{sub:4} describing these computations in detail. 
\begin{algorithm}
\caption{Local computation of controller signals}\label{alg:full}
\begin{algorithmic}[1]
\For{Each node $i = 1, \ldots, N$}
    \State Input $\tPhisf_{w, \alpha}(:, i)$, $\tPhisf_{w, \zeta}(:, i)$, $\tPhifc_{ew}(:, i)$, $\tPhifc_{ev}(:, i)$
    \State Initialize $\beta_i(0) \gets 0$, $w_i(t) \gets 0$, $v_i(t) \gets 0$
\EndFor
\For{$t = 0, 1,  \ldots$}
\For{Each node $i = \{1, \ldots, N\}$} \Comment{parallel}
    \State Observe $y_i(t)$
    \State Receive $\beta_j(t)$ and $y_j(t)$ from $j \in \Nin^d(i)$ \label{line:receive}
    \State subroutine1()
    \State Receive $\widehat w_j(t)$ and $\widehat v_j(t)$ from $j \in \Nin^1(i)$ \label{line:wv}
    \State subroutine2()
    \State Receive $\widehat \alpha^{(\Nout^d(j))}(t+1)$, $\widehat \zeta^{(\Nout^d(j))}(t+1)$, $\widehat \gamma^{(\Nout^d(j))}(t+1)$ and $\widehat \theta^{(\Nout^d(j))}(t)$ from $j \in \Nin^d(i)$. \label{line:greek}
    \State subroutine3()
    \State Receive $\alpha_j(t + 1)$ and $\gamma_j(t + 1)$ from $j\in \Nin^1(i)$ \label{line:ag}
    \State subroutine4()
    \State Apply $u_i(t)$
\EndFor
\EndFor
\end{algorithmic}
\end{algorithm}

Control signal computation at subsystem $i$ begins by receiving the measurements from neighbors $j$ at most $d$ steps away (line~\ref{line:receive}) and computing the $i$th element of the internal signals $\hat v(t + 1)$ and $\hat w(t + 1)$ via Subroutine~\ref{sub:1}, which is illustrated in Fig.~\ref{sub:1}. Here the function $\text{step}(G, u)$ means that the internal dynamics of the system $G$ is propagated one time-step with the input $u$.

\begin{Subroutine}
\caption{Compute $\widehat w_i(t + 1)$ and $\widehat v_i(t + 1)$}\label{sub:1}
\begin{algorithmic}
    \State Receive $\beta_j(t)$ and $y_j(t)$ from $j \in \Nin^d(i)$
    \State $\beta_{[\Nin^d(i)]}(t) \gets \text{vec}(\beta_{j_1}(t), \ldots, \beta_{j_m}(t))$
    \State $y_{[\Nin^d(i)]}(t) \gets \text{vec}(y_{j_1}(t), \ldots, y_{j_m}(t))$
    \State $\widehat w_i(t + 1) \gets \text{step}(z\tPhifc_{ew}(i, :), \beta_{[\Nin^d(i)]}(t))$
    \State $\widehat v_i(t + 1) \gets \text{step}(z\tPhifc_{ev}(i, :), y_{[\Nin^d(i)]}(t))$
\end{algorithmic}
\end{Subroutine}

\begin{figure}
    \centering
    \scalebox{1}{\begin{tikzpicture}[node distance = 4.5em, line width=0.5mm, inner ysep=6pt, >=latex]
    \node[sig] (yi) {$\mathbf y_i$};
    \node[sig, above = 2em of yi] (y1) {$\mathbf y_{j_1}$};
    \node[sig, below = 2em of yi] (ym) {$\mathbf y_{j_m}$};
    
    \node at ($(y1.south)!0.4!(yi.north)$) {$\vdots$};
    \node at ($(yi.south)!0.4!(ym.north)$) {$\vdots$};
    
    \node[right of=y1] (muxtop) {};
    \node[right of=ym] (muxbottom) {};
    \node (muxmid) at ($(muxtop)!0.5!(muxbottom)$) {};
    \node[block, right=4em of muxmid] (Phifc) {$\tPhifc_{ev}(i, :)$};
    \node[sig, right=4em of Phifc] (vi) {$\widehat{\mathbf{v}}_i$};
    
    \draw[-, line width=1mm] (muxtop.north) -- (muxbottom.south);
    \draw[->] (y1) -- (muxtop);
    \draw[->] (yi) -- (muxmid);
    \draw[->] (ym) -- (muxbottom);
    \draw[->] (muxmid) --  node[below] {$\mathbf y_{[\Nin^d(i)]}$} (Phifc);
    \draw[->] (Phifc) -- (vi);
    
\end{tikzpicture}}
    \caption{Illustration of subroutine 1. The computation of $\widehat w$ is similar.}
        \label{fig:sub1}
\end{figure}
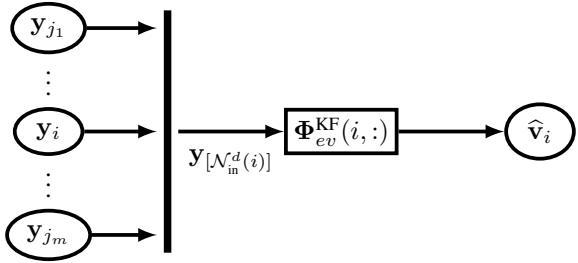

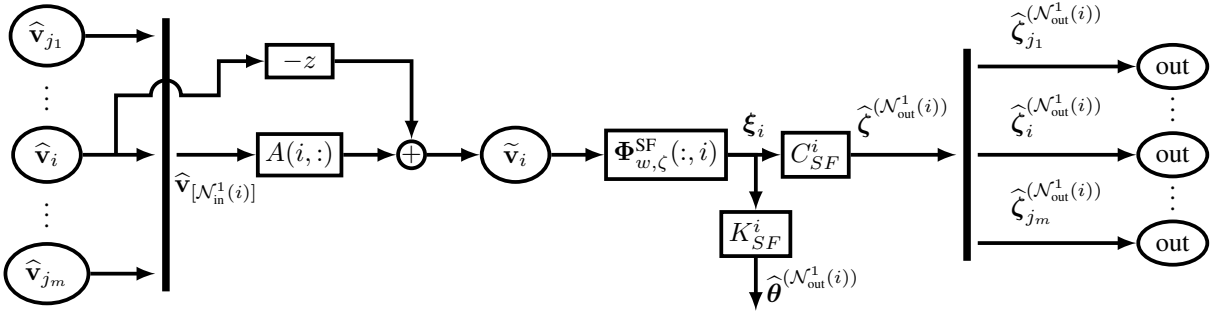
\begin{figure*}
    \centering
    \scalebox{1}{\begin{tikzpicture}[node distance = 4.5em, line width=0.5mm, inner ysep=6pt, >=latex]
    \node[sig] (vhati) {$\widehat{\mathbf{v}}_i$};
    \node[sig, above of=vhati] (vhat1) {$\widehat{\mathbf{v}}_{j_1}$};
    \node[sig, below of=vhati] (vhatm) {$\widehat{\mathbf{v}}_{j_m}$};
    
    \node at ($(vhat1.south)!0.4!(vhati.north)$) {$\vdots$};
    \node at ($(vhati.south)!0.4!(vhatm.north)$) {$\vdots$};
    
    \node[right of=vhat1] (muxtop) {};
    \node[right of=vhatm] (muxbottom) {};
    \node (muxmid) at ($(muxtop)!0.5!(muxbottom)$) {};
    \node[block, right=3em of muxmid] (A) {$A(i, :)$};
    \node[block, above=2em of A] (shift) {$-z$};
    \node[sum, right=2em of A] (sum) {$+$};
    \node[sig, right=2em of sum] (vtildei) {$\widetilde{\mathbf{v}}_i$};
    
    \draw[-, line width=1mm] (muxtop.north) -- (muxbottom.south);
    \draw[->] (vhat1) -- (muxtop);
    \draw[->] (vhati) -- (muxmid);
    \draw[->] (vhatm) -- (muxbottom);
    \draw[->] (muxmid) -- node[below] {$\widehat{\mathbf{v}}_{[\Nin^1(i)]}$} (A);
    \draw[->] (A) -- (sum);
    \draw[->] (sum) -- (vtildei);
    
    \draw[->] ($(vhati.east)!0.4!(muxmid)$) {
    [shift only] |- ($(muxmid)!0.5!(muxtop)+(-2mm, 0)$) arc(180:0:2mm)  
        -- ($(muxmid)!0.5!(muxtop)+(2em, 0)$)  
        |- (shift)
    };
    \draw[->] (shift) -| (sum);

    \node[block, right = 2em of vtildei] (Phisf) {$\tPhisf_{w, \zeta}(:, i)$};
    \node[block, right = 2em of Phisf] (C) {$C_{SF}^i$};
    \node[right = 4em of C] (muxmid1) {};
    \node[above = 2em of muxmid1] (muxtop1) {};
    \node[below = 2em of muxmid1] (muxbottom1) {};
    
    \node[sig, right=6em of muxtop1] (zetaout1) {out};
    \node[sig, right=6em of muxmid1] (zetaouti) {out};
    \node[sig, right=6em of muxbottom1] (zetaoutm) {out};
    
    \node at ($(zetaout1)!0.4!(zetaouti)$) {$\vdots$};
    \node at ($(zetaouti)!0.4!(zetaoutm)$) {$\vdots$};
    
    \draw[->] (vtildei) -- (Phisf);
    \draw[->] (Phisf) -- node[above] {$\boldsymbol \xi_i$} (C);
    \draw[->] (C) -- node[above] {$\widehat \bzeta^{(\Nout^1(i))}$} (muxmid1);
    \draw[-, line width=1mm] (muxtop1.north) -- (muxbottom1.south);
    \draw[->] (muxtop1) -- node[above] {$\widehat\bzeta^{(\Nout^1(i))}_{j_1}$} (zetaout1);
    \draw[->] (muxmid1) -- node[above] {$\widehat\bzeta^{(\Nout^1(i))}_{i}$} (zetaouti);
    \draw[->] (muxbottom1) -- node[above] {$\widehat\bzeta^{(\Nout^1(i))}_{j_m}$} (zetaoutm);
    
    \node[block] (K) at ([yshift=-3em]$(Phisf)!0.6!(C)$) {$K^i_{SF}$};
    \node[below = 2em of K] (vdots){};
    \draw[->] (Phisf) -| (K);
    \draw[->] (K) -- node[right] {$\widehat \btheta^{(\Nout^1(i))}$} (vdots);
    
\end{tikzpicture}}
    \caption{Illustration of subroutine 2.}
    \label{fig:sub2}
\end{figure*}

In the second stage, the node receives $\widehat v_j(t)$ and $\widehat w_j(t)$ from its closest neighbors (line~\ref{line:wv}) and computes the outgoing components of \eqref{eq:greek_letters}.
The computations are outlined in Subroutine~\ref{sub:2} and illustrated in Fig.~\ref{fig:sub2}.

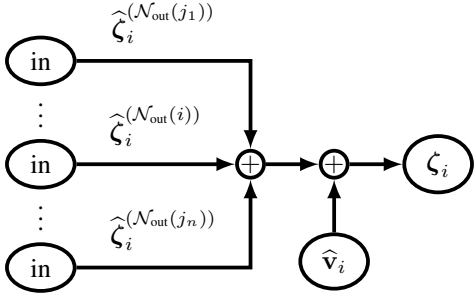
\begin{figure}
    \centering
    \scalebox{1}{\begin{tikzpicture}[node distance = 4.5em, line width=0.5mm, inner ysep=6pt, >=latex]
    \node[sig] (zetaini) {in};
    \node[sig, above = 2em of zetaini] (zetain1) {in};
    \node[sig, below = 2em of zetaini] (zetainn) {in};

    \node at ($(zetain1)!0.45!(zetaini)$) {$\vdots$};
    \node at ($(zetaini)!0.45!(zetainn)$) {$\vdots$};
    
    \node[sum, right=6em of zetaini] (sum1) {$+$};
    \node[sum, right=2em of sum1] (sum2) {$+$};
    \node[sig, below = 2em of sum2] (vhati) {$\widehat{\mathbf{v}}_i$};
    \node[sig, right=2em of sum2] (zetai) {$\boldsymbol{\zeta}_i$};
    
    \draw[->] (zetain1) -| node[above, pos = 0.25] {$\widehat\bzeta^{(\Nout(j_1))}_{i}$} (sum1);
    \draw[->] (zetainn) -| node[above, pos = 0.25] {$\widehat\bzeta^{(\Nout(j_n))}_{i}$} (sum1);
    \draw[->] (zetaini) -- node[above] {$\widehat\bzeta^{(\Nout(i))}_{i}$} (sum1);
    \draw[->] (sum1) -- (sum2);
    \draw[->] (vhati) -- (sum2);
    \draw[->] (sum2) -- (zetai);    
\end{tikzpicture}}
    \caption{Illustration of subroutine 3.}
    \label{fig:sub3}
\end{figure}

\begin{Subroutine}
\caption{Compute the outgoing components of \eqref{eq:greek_letters}}\label{sub:2}
\begin{algorithmic}
    \State $\widehat w_{[\Nin^1(i)]}(t) \gets \text{vec}(\widehat w_{j_1}(t), \ldots, \widehat w_{j_m}(t))$
    \State $\widehat v_{[\Nin^1(i)]}(t) \gets \text{vec}(\widehat v_{j_1}(t), \ldots, \widehat v_{j_m}(t)) $  \vspace{.2em}
    \State $\widetilde w_i(t) = \beta_i(t) +A\widehat w_{[\Nin^1(i)]}(t) - \widehat w_i(t+1)$ 
    \State $\widetilde v_i(t) = A\widehat v_{[\Nin^1(i)]}(t) - \widehat v_i(t+1)$
    \State $\lambda_i(t + 1) \gets \text{step}(\tPhisf_{w, \alpha}(:, i), e_i\widetilde w_i(t))$
    \State $\xi_i(t + 1)\gets \text{step}(\tPhisf_{w, \zeta}(:, i), e_i\widetilde v_i(t))$
    \State $\widehat \alpha^{(\Nout^d(j))}(t+1)  \gets C_{SF}^i\lambda_i(t + 1)$
    \State $\widehat \zeta^{(\Nout^d(j))}(t+1)  \gets C_{SF}^i \xi_i(t + 1)$
    \State $\widehat \gamma^{(\Nout^d(j))}(t + 1) \gets K^i_{SF}\lambda_i(t + 1)$
    \State $\widehat \theta^{(\Nout^d(j))}(t) \gets K^i_{SF}\xi_i(t + 1)$
\end{algorithmic}
\end{Subroutine}
In the third stage, which is demonstrated in Fig.~\ref{fig:sub3}, the node receives the components pertaining to its element of the signals in \eqref{eq:certainty_equivalence} from other nodes a distance at most $d$ steps away with delayed information (line~\ref{line:greek}) and sums them to compute the $i$th element of each esignal in $\eqref{eq:greek_letters}$.
This step is described in Subroutine~\ref{sub:3}.

\begin{Subroutine}
\caption{Compute the local elements of \eqref{eq:greek_letters}}\label{sub:3}
\begin{algorithmic}
    \State $\alpha_i(t+1) \gets \widehat w_i(t+1) + \sum_j\widehat\alpha^{(\Nout^1(j))}_i(t+1)$
    \State $\gamma_i(t + 1) \gets \sum_j \widehat\gamma^{(\Nout^1(j))}_i(t+1)$
    \State $\zeta_i(t + 1) \gets \widehat v_i(t+1) + \sum_j \widehat\zeta^{(\Nout^1(j))}_i(t + 1)$
    \State $\theta_i(t) \gets \sum_j\widehat\theta^{(\Nout^1(j))}_i(t)$
\end{algorithmic}
\end{Subroutine}

In the final stage (Subroutine~\ref{sub:4}) the node receives $\alpha_j(t + 1)$ and $\gamma_j(t + 1)$ from its closest neighbors (line~\ref{line:ag}) and computes $\beta_i(t + 1)$ and $u_i(t)$. We conclude that the node has now received information from nodes at most $2d + 2$ steps away.

\begin{Subroutine}
\caption{Compute $u_i(t)$ and $\beta_i(t + 1)$}\label{sub:4}
\begin{algorithmic}
    \State $\alpha_{[\Nin^1(i)]}(t + 1) \gets \text{vec}\left(\alpha_{j_1}(t + 1), \ldots, \widehat \alpha_{j_m}(t + 1)\right)$
    \State $\gamma_{[\Nin^1(i)]}(t + 1) \gets \text{vec}\left(\gamma_{j_1}(t + 1), \ldots, \widehat \gamma_{j_m}(t + 1)\right)$ \vspace{-.75em}
    \begin{multline*}
        \beta_i(t + 1) \gets -A(i, :)\alpha_{[\Nin^1(i)]}(t + 1)\\ - B(i, :)\gamma_{[\Nin^1(i)]}(t + 1) - \zeta_i(t + 1)
    \end{multline*}%
    \vspace{-1.5em}
    \State $u_i(t) \gets \gamma_i(t + 1) + \theta_i(t)$
\end{algorithmic}
\end{Subroutine}
We summarize the stability properties of Algorithm~\ref{alg:full} in the following theorem:
\begin{theorem}
    Algorithm~\ref{alg:full} with $\tPhisf_w$ as in \eqref{eq:phw}, $\tPhifc_{ew}$ and $\tPhifc_{ev}$ as in Section~\ref{sec:delay} internally stabilizes system~\eqref{eq:dynamics}. Moreover if $\tPhisf_w$, $\tPhifc_{ew}$ and $\tPhifc_{ev}$ are $d$-localized, the closed-loop is at most $2d + 2$-localized.
\end{theorem}
\begin{proof}
By Theorem~\ref{thm:observer_feedback} the closed-loop maps satisfy \eqref{eq:feasibility_x} and \eqref{eq:feasibility_u}. 
Concatenating $\by_i$, $\bbeta_i$ and $\bu_i$ we get precisely the signals in Fig.~\ref{fig:sls_of} which is internally stable~\cite{anderson2019system}, we need to show that the closed-loop is internally stable for perturbations entering in the intermediate steps outlined in Subroutines~\ref{sub:1}--\ref{sub:4}.
Note that a perturbation entering at any of the intermediate signals can be modeled as a disturbance entering as $\boldsymbol \delta_x,\ \boldsymbol \delta_y$ or $\boldsymbol \delta_\beta$ pre-filtered through a stable linear system.
Similarly, probing any of the internal signals can be represented as probing $\by,\ \bu$ or $\bbeta$ post-filtered through a stable system.
We conclude Algorithm~\ref{alg:full} is internally stable in feedback with System~\ref{eq:dynamics}.
Finally, as $d$-localization is closed under addition, and composition of a $d$- and a $k$-localized operator is at most $d+k$-localized, \eqref{eq:certainty_equivalence} implies that the closed loop is at most $2d + 2$-localized.
\end{proof}

\section{Numerical simulations}
\label{sec:simulation}
\begin{figure}
 \centering
 \includegraphics[clip,trim=1.5cm 9.2cm 1.5cm 9.2cm,scale = 0.45]{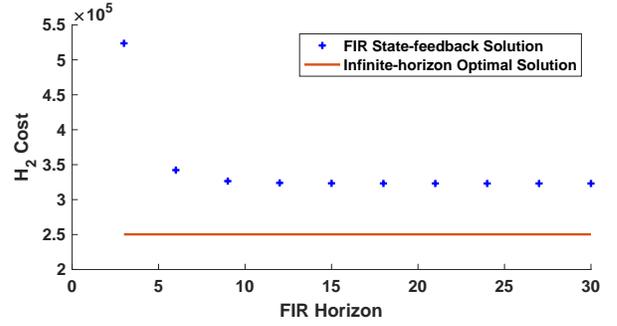}
 \caption{The infinite-horizon SLS solution achieves the optimal cost.}
 \label{fig:optimality}
\end{figure}

Consider a bi-directional scalar chain network parameterized by $\alpha$ and $\rho$: 

\begin{equation*}
        x^i[t+1] = \rho (1-2\alpha ) x^i[t] + \rho \alpha \sum_{j \in \{ i\pm 1\}} x^j[t] + u^i[t] + w^i[t]
    \end{equation*}

where $\alpha$ {is a coupling constant} and $\rho$ is the spectral radius of the global state-transition matrix $A$, with $\rho\geq1$ being unstable. We first verify the optimality of the infinite-horizon state-feedback solution given in Section \ref{sec:delay}.  In this simulation, we choose the number of scalar subsystems to be 15, $\alpha = 0.6$ and $\rho = 1$. For the quadratic cost matrices, we let $Q = I$ and $R = 300\cdot I$. For SLCs, we let the delayed localization parameter be $d=3$. The result is shown in Figure \ref{fig:optimality}, where the optimality of our approach is clear. Due to the high penalty on the control actions, the performance degradation under FIR approximation can be significant.

\begin{figure}
 \centering
 \includegraphics[clip,trim=1.5cm 9.2cm 1.5cm 9.2cm,scale = 0.45]{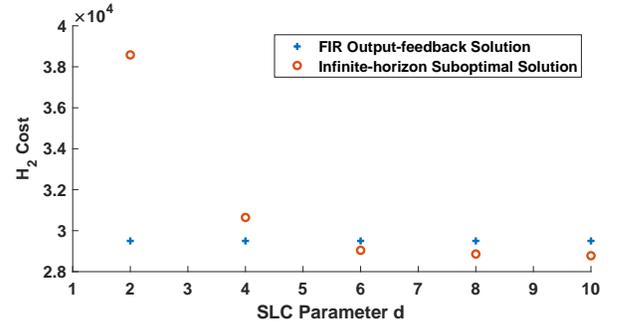}
 \caption{The proposed infinite-horizon suboptimal solution to the output-feedback SLS problem versus the FIR output-feedback solution numerically computed for \eqref{eq:of-original} for varying SLC delayed localization parameter $d$.}
 \label{fig:varying_d}
\end{figure}

Next, we investigate the optimality gap between the suboptimal infinite-horizon output-feedback solution proposed in this work against the FIR output-feedback solution computed numerically with a fixed FIR horizon of 20. We let all other parameters remain the same as before, and change the number of subsystems to $10$ in this simulation. First, we study how the $d$ delayed localization parameter influence the optimality gap. This is illustrated in Figure \ref{fig:varying_d}. As expected, the more localized the output-feedback problem is, the bigger the optimality gap is between the constructed solution using separation principle and the direct FIR output-feedback solution. As the delayed localization pattern becomes more global, the proposed output-feedback solution becomes more optimal. When the delayed localization SLCs become non-binding (for $d \geq 6$), we see that the proposed infinite-horizon output-feedback solution actually becomes optimal and achieves lower cost than the FIR solution. This is due to the separation principle of centralized LQG. 

\begin{figure}
 \centering
 \includegraphics[clip,trim=1.5cm 9.2cm 1.5cm 9.2cm,scale = 0.48]{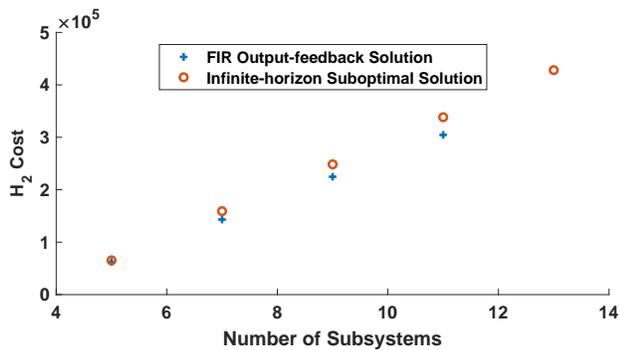}
 \caption{The proposed suboptimal solution to the output-feedback SLS problem versus the FIR output-feedback solution numerically computed for \eqref{eq:of-original} for varying number of subsystems in the network.}
 \label{fig:varying_N}
\end{figure}

Next, we investigate how the optimality gap grows with the number of subsystems in the network. Here we have fixed the delayed localization parameter to be $d=3$. As can be seen in Figure \ref{fig:varying_N}, we observe that the optimality gap grows apparently linearly in the number of subsystems. However, we highlight the numerical efficiency and stability of our approach despite the suboptimality. When the number of subsystems exceeds 12 with FIR horizon of 20, the FIR solution solved in MATLAB using CVX renders NaN due to numerical instability (total of 11520 variables). 



\bibliographystyle{IEEEtran}
\bibliography{references}

\end{document}